\documentclass[pra,aps,amsmath,amssymb,showkeys,showpacs,eqsecnum]{revtex4-2}
\usepackage{graphicx}

\newcommand{\hh}{{\mathcal{H}}}
\newcommand{\lnp}{{\mathcal{L}}}
\newcommand{\lsp}{{\mathcal{L}}_{+}}

\newcommand{\en}{{\mathbb{E}}}
\newcommand{\qn}{{\mathbb{Q}}}
\newcommand{\psym}{{\mathsf{\Pi}}_{\mathrm{sym}}}
\newcommand{\asym}{{\mathsf{\Pi}}_{\mathrm{asym}}}
\newcommand{\pen}{\openone}
\newcommand{\bro}{\boldsymbol{\rho}}
\newcommand{\bogm}{\boldsymbol{\varpi}}
\newcommand{\bau}{\boldsymbol{\tau}}
\newcommand{\me}{{\mathsf{E}}}
\newcommand{\mg}{{\mathsf{G}}}
\newcommand{\am}{{\mathsf{A}}}
\newcommand{\pq}{{\mathsf{P}}}
\newcommand{\qp}{{\mathsf{Q}}}
\newcommand{\tr}{\mathrm{tr}}
\newcommand{\iu}{\mathtt{i}}
\newcommand{\clb}{{\mathcal{B}}}
\newcommand{\cle}{{\mathcal{E}}}
\newcommand{\clp}{{\mathcal{P}}}
\newcommand{\clq}{{\mathcal{Q}}}
\newcommand{\fw}{\tilde{\mathsf{W}}}
\newcommand{\con}{\mathcal{C}}

\begin{document}
\clearpage
\preprint{}

\title{Uncertainty relations for quantum measurements from generalized equiangular tight frames}

\author{Alexey E. Rastegin}

\affiliation{Department of Theoretical Physics, Irkutsk State University, K. Marx St. 1, Irkutsk 664003, Russia}

\begin{abstract}
The current study aims to examine uncertainty relations for
measurements from generalized equiangular tight frames.
Informationally overcomplete measurements are a valuable tool in
quantum information processing, including tomography and state
estimation. The maximal sets of mutually unbiased bases are the most
common case of such measurements. The existence of $d+1$ mutually
unbiased bases is proved for $d$ being a prime power. More general
classes of informationally overcomplete measurements have been
proposed for various purposes. Measurements of interest are
typically characterized by some inner structure maintaining the
required properties. It leads to restrictions imposed on generated
probabilities. To apply the considered measurements, these
restrictions should be converted into information-theoretic terms.
It is interesting that certain restrictions hold irrespectively to
overcompleteness. To describe the amount of uncertainty
quantitatively, we use the Tsallis and R\'{e}nyi entropies as well
as probabilities of separate outcomes. The obtained results are
based on estimation of the index of coincidence. The derived
relations are briefly exemplified.
\end{abstract}


\maketitle

\pagenumbering{arabic}
\setcounter{page}{1}

\section{Introduction}\label{sec1}

Quantum technologies of information processing are currently the
subject of intensive studies \cite{nielsen}, though complete
awareness of their role belongs to the future. The final stage of
various protocols for quantum information processing involves
measurements required for obtaining information about resulting
quantum states. Quantum measurements are represented by positive
operator-valued measures (POVMs) \cite{watrous}. Each of them
consists of positive semidefinite operators that sum up to the
identity operator. To ensure convenient dealing with, the utilized
measurements are built to have good properties. A measurement is
informationally complete, when every possible state is uniquely
determined by the measurement statistics \cite{prug}. Complete sets
of mutually unbiased bases are very important example
\cite{schwinger,ivano81,fields89}. Symmetric informationally
complete measurements (SIC-POVMs) have many nice properties to use
\cite{rbksc04,fhs2017}. Fisher-symmetric informationally complete
measurements and their applications were discussed in
\cite{gross,zhu}. Experimental implementation of symmetric
informationally complete measurements were described in
\cite{meden,padua,bent,rosset,smania}.

In fact, the problem of building desired structures in Hilbert
spaces is often hard to resolve \cite{fhs2017,abfg19}. It is known
that $d+1$ mutually unbiased bases in dimension $d$ exist for $d$
being a prime power. But the maximal number of mutually unbiased
bases remains unknown even for $d=6$ \cite{bz10}. There are reasons
to believe that SIC-POVMs exist for all $d$, and this conjecture is
due to Zauner \cite{zauner11}. Here, there are ways to weak some of
the imposed restrictions. Dealing with rank-one measurement
operators, we may reduce the number of outcomes. In this regard,
equiangular tight frames are shown to deserve more attention than
they have obtained \cite{rastfram}. Equiangular measurements are a
special case of more general concepts such as equioverlapping
measurements \cite{feng22,feng24} and the so-called $(N,M)$-POVMs
\cite{smfd,sent}. Another way is to allow measurement operators of
arbitrary rank. General symmetric informationally complete
measurements were built \cite{gour2014,kimura} and applied
\cite{lysmf}. Mutually unbiased measurements instead of bases were
proposed in \cite{kalev2014}. Recently, generalized symmetric
measurements \cite{symm} and generalized equiangular ones
\cite{getf} have been studied.

The Heisenberg uncertainty principle \cite{heisenberg} is widely
recognized as a fundamental concept that the observer perturbs what
he observes. Since the first formal derivations of Kennard
\cite{kennard} and Robertson \cite{robertson} appeared, many
approaches and scenarios were addressed \cite{bpl07}. For
finite-dimensional systems, entropic uncertainty relations
\cite{deutsch,maass88} give a useful alternative to Robertson's
formulation. The well-known Maassen--Uffink uncertainty relation
\cite{maass88} was originally conjectured by Kraus \cite{kraus87}.
This relation becomes a tool in building feasible schemes to detect
nonclassical correlations such as entanglement \cite{olew,olew1} and
steerability \cite{brun18,cug18,cugen}. Entropic uncertainty
relations have also emerged in the security analysis of quantum
cryptographic protocols \cite{koashi,molm09b,ng12}. Entropic
formulation gives a natural way to address uncertainties in the
presence of quantum memory \cite{bccrrr10}. For more results on
entropic uncertainty relations, see the reviews
\cite{ww10,cbtw17,cerf} and references therein. For measurements
with a special inner structure, the Maassen--Uffink uncertainty
relation can be improved \cite{rastmubs,rastnot,rastopn,feiqip}.

It is important to examine properties of specially built quantum
measurements as completely as possible, even if the final opinion on
their use is now unformed. The aim of this study is to derive
uncertainty relations for quantum measurements from generalized
equiangular tight frames. In contrast to the treatment of
\cite{getf}, these measurements are not assumed to be overcomplete.
The new relations are expressed in terms of the R\'{e}nyi
\cite{renyi61} and Tsallis entropies \cite{tsallis} as well as sums
of probabilities. Some of the used inequalities are based on
considering information diagrams as described in
\cite{rastrid,raplq}. The paper is organized as follows. Section
\ref{sec2} introduces the preliminary material concerning
generalized equiangular measurements and generalized symmetric ones.
Some properties of the Tsallis and R\'{e}nyi entropies are recalled
as well. New uncertainty relations for the considered measurements
are presented in Sec. \ref{sec3}. A utility of the obtained
inequalities is briefly exemplified. Section \ref{sec4} concludes
the paper with a summary of the results. Appendix \ref{incva} deals
with a statement concerning indices of coincidence.

\section{Preliminaries}\label{sec2}

This section aims to recall the required material concerning quantum
measurements derived from generalized equiangular tight frames. It
also considers entropies that will be used to characterize
uncertainties in quantum measurements of interest.

\subsection{Generalized equiangular measurements and conical $2$-designs}\label{sbc21}

Basic reasons to put informationally overcomplete measurements are
discussed in \cite{getf}. We recall only very necessary facts about
them. Let $\hh_{d}$ be $d$-dimensional Hilbert space. By
$\lnp(\hh_{d})$ and $\lsp(\hh_{d})$, we respectively mean the space
of linear operators on $\hh_{d}$ and the set of positive
semidefinite ones. The state of a quantum system is described by
density matrix $\bro\in\lsp(\hh_{d})$ normalized as $\tr(\bro)=1$.
In general, any quantum measurement is represented as a set
$\cle=\bigl\{\me_{j}\bigr\}_{j=1}^{N}$ with
$\me_{j}\in\lsp(\hh_{d})$ such that the completeness relation
holds \cite{bcm96}
\begin{equation}
\sum_{j=1}^{N}\me_{j}=\pen_{d}
\, . \label{compr}
\end{equation}
For the pre-measurement state $\bro$, the probability of $j$-th
outcome is equal to $\tr(\me_{j}\bro)$. It is very important that
the number of different outcomes can exceed the dimensionality.

A collection of $M$ generalized equiangular lines
$\clp_{\mu}=\bigl\{\pq_{\mu,j}\bigr\}_{j=1}^{N_{\mu}}$ with
$\pq_{\mu,j}\in\lsp(\hh_{d})$ is a generalized equiangular
measurement if the following items hold \cite{getf}:
\begin{align}
\sum_{j=1}^{N_{\mu}}\pq_{\mu,j}&=\gamma_{\mu}\pen_{d}
\, , \label{cn0}\\
\sum_{\mu=1}^{M}\gamma_{\mu}&=1
\, , \label{cn01}\\
\tr(\pq_{\mu,j})&=a_{\mu}
\, , \label{cn1}\\
\tr(\pq_{\mu,j}^{2})&
=b_{\mu}\tr(\pq_{\mu,j})^{2}
\, , \label{cn2}\\
\tr(\pq_{\mu,i}\pq_{\mu,j})&=c_{\mu}\tr(\pq_{\mu,i})\,\tr(\pq_{\mu,j})
\qquad (i\neq{j})
\, , \label{cn3}\\
\tr(\pq_{\mu,i}\pq_{\nu,j})&=f\,\tr(\pq_{\mu,i})\,\tr(\pq_{\nu,j})
\qquad (\mu\neq\nu)
\, . \label{cn4}
\end{align}
From the above definition, it follows that $f=d^{-1}$,
\begin{equation}
a_{\mu}=\frac{\gamma_{\mu}d}{N_{\mu}}
\ , \qquad
c_{\mu}=\frac{N_{\mu}f-b_{\mu}}{N_{\mu}-1}
\ . \label{coels}
\end{equation}
A single equiangular measurement with rank-one elements is a
particular case for $M=\gamma=b=1$. The generalized equiangular
measurement can also be interpreted as a set of $M$ POVMs so that
$\mu$-th POVM has elements $\gamma_{\mu}^{-1}\,\pq_{\mu,j}$. Due to
this correspondence, a generalized equiangular measurement relates
to some generalized symmetric measurement \cite{getf}.

The generalized symmetric measurements were introduced as
collections of POVMs that are not equinumerous \cite{symm}. A
generalized equiangular measurement can be constructed from elements
of generalized symmetric measurements. One uses \cite{symm} a set
of $M$ POVMs $\cle_{\mu}=\bigl\{\me_{\mu,j}\bigr\}_{j=1}^{N_{\mu}}$
satisfying additional symmetry conditions:
\begin{align}
\tr(\me_{\mu,j})&=w_{\mu}
\, , \label{sm1}\\
\tr(\me_{\mu,j}^{2})&=x_{\mu}
\, , \label{sm2}\\
\tr(\me_{\mu,i}\me_{\mu,j})&=y_{\mu}
\qquad(i\neq{j})
\, , \label{sm3}\\
\tr(\me_{\mu,i}\me_{\nu,j})&=z_{\mu\nu}
\qquad(\mu\neq\nu)
\, . \label{sm4}
\end{align}
For informationally complete measurements, one has
\begin{equation}
\sum_{\mu=1}^{M}N_{\mu}=d^{2}+M-1
\, . \label{eqmn}
\end{equation}
Let positive numbers $\gamma_{\mu}$ obey (\ref{cn01}) and $M$ POVMs
$\cle_{\mu}=\bigl\{\me_{\mu,j}\bigr\}_{j=1}^{N_{\mu}}$ satisfy
(\ref{sm1})--(\ref{sm4}). Then operators of the form
\begin{equation}
\pq_{\mu,j}=\gamma_{\mu}\,\me_{\mu,j}
 \label{linr}
\end{equation}
compose a generalized equiangular measurement \cite{getf}. The
correspondence between the coefficients reads as
\begin{equation}
a_{\mu}=\gamma_{\mu}w_{\mu}
\, , \qquad
b_{\mu}=\frac{x_{\mu}}{w_{\mu}^{2}}
\ , \qquad
c_{\mu}=\frac{y_{\mu}}{w_{\mu}^{2}}
\ , \qquad
f=\frac{z_{\mu\nu}}{w_{\mu}w_{\nu}}
\ . \label{abcf}
\end{equation}
The obtained POVM is informationally complete, if it is constructed
from informationally complete set of elements $\me_{\mu,j}$ and
$\gamma_{\mu}>0$ for all $\mu=1,\ldots,M$ \cite{getf}. Despite of
simplicity of the relation (\ref{linr}), this construction has some
nice features. For example, one can manipulate the coefficients so
that both the traces $\tr(\pq_{\mu,j}^{2})$ and
$\tr(\pq_{\mu,i}\pq_{\mu,j})$ become $\mu$-independent \cite{getf}.

Under additional conditions on the coefficients, a generalized
equiangular measurement is a conical $2$-design \cite{getf}.
Projective and conical designs have found use in various questions
of quantum information theory. Let
$\bigl\{|\phi_{j}\rangle\bigr\}_{j=1}^{K}$ be a set of $K$ unit
vectors of $\hh_{d}$. There are several equivalent definitions, when
these vectors form a projective $2$-design \cite{zauner11,scott}.
One of them reads as
\begin{equation}
\frac{1}{K}\,\sum_{j=1}^{K} |\phi_{j}\rangle\langle\phi_{j}|\otimes|\phi_{j}\rangle\langle\phi_{j}|
=\frac{2}{d^{2}+d}\>\psym
\, , \label{def1}
\end{equation}
where $\psym$ is the orthogonal projection on the symmetric subspace
of $\hh_{d}\otimes\hh_{d}$. Conical designs are generalizations of
the above concept to positive operators \cite{gapp16a,gapp16b}. We
introduce the unitary swap operator $\fw$ on $\hh_{d}\otimes\hh_{d}$
which takes $|\psi\rangle\otimes|\varphi\rangle$ to
$|\varphi\rangle\otimes|\psi\rangle$. The set
$\bigl\{\am_{j}\bigr\}_{j=1}^{K}$ with $\am_{j}\in\lsp(\hh_{d})$ is
a conical $2$-design, if and only if
\begin{equation}
\sum_{j=1}^{K} \am_{j}\otimes\am_{j}=\kappa_{\mathrm{s}}\psym+\kappa_{\mathrm{a}}\asym=\kappa_{+}\pen_{d}\otimes\pen_{d}+\kappa_{-}\fw
\, , \label{con2d}
\end{equation}
where $\kappa_{+}\geq\kappa_{-}>0$. It can be noticed that
$2\kappa_{\pm}=\kappa_{\mathrm{s}}\pm\kappa_{\mathrm{a}}$. To obtain
a conical $2$-design from equiangular lines, the coefficients should
satisfy \cite{getf}
\begin{equation}
a_{\mu}^{2}(b_{\mu}-c_{\mu})=S
 \label{asmin}
\end{equation}
for all $\mu$. Then the generalized equiangular measurement is a
conical 2-design with
\begin{equation}
\kappa_{+}=\frac{\sigma-S}{d}
\, , \qquad
\kappa_{-}=S
\, , \qquad
\sigma=\sum_{\mu=1}^{M} a_{\mu}\gamma_{\mu}
\, . \label{kaps}
\end{equation}
There exist conical $2$-designs even though they are constructed
from the generalized symmetric measurements that are not conical
$2$-designs \cite{getf}. This property is useful in producing new
conical $2$-designs. Unitary designs are also considered as a
powerful tool in quantum information science \cite{cirac18,chart19}.
It turned out that generated probabilities satisfy certain
conditions that lead to various restrictions including uncertainty
relations. To express the amount of uncertainties, the Tsallis and
R\'{e}nyi entropies will be used in this paper.

\subsection{Generalized entropies and the index of coincidence}\label{sbc22}

The index of coincidence is used in various questions of information
theory \cite{harr2001}. Let $P=(p_{1},\ldots,p_{N})$ be a
probability distribution with the index running over the set
$\{1,\ldots,N\}$. The index of coincidence reads as
\begin{equation}
I(P)=\sum_{j=1}^{N} p_{j}^{2}
\, . \label{icon}
\end{equation}
Indices with larger degrees of probabilities have also found use
\cite{harr2001}. Applications of such indices to derive uncertainty
relations were given in \cite{guhne20,rast20d,rastsiam}. For
$\alpha>0$, the Tsallis $\alpha$-entropy is expressed as
\cite{tsallis}
\begin{equation}
H_{\alpha}(P)=\frac{1}{1-\alpha}\,\biggl(\,\sum_{j=1}^{N} p_{j}^{\alpha}-1\biggr)
=\sum_{j=1}^{N}p_{j}\ln_{\alpha}\biggl(\frac{1}{p_{j}}\biggr)
 . \label{tsed}
\end{equation}
The right-hand side of (\ref{tsed}) uses the $\alpha$-logarithm of
positive variable,
\begin{equation}
\ln_{\alpha}(X)=
\begin{cases}
 \frac{X^{1-\alpha}-1}{1-\alpha} \>, & \text{if}\ 0<\alpha\neq1 \, , \\
 \ln{X} \, , & \text{if}\ \alpha=1 \, .
\end{cases}
\label{lnal}
\end{equation}
Substituting the probabilities $\tr(\me_{j}\bro)$ into (\ref{tsed})
gives the entropy $H_{\alpha}(\cle;\bro)$. Some results of the
current study deal with the R\'{e}nyi $\alpha$-entropy defined as
\cite{renyi61}
\begin{equation}
R_{\alpha}(P)=\frac{1}{1-\alpha}\,\ln\biggl(\,\sum_{j=1}^{N} p_{j}^{\alpha}\biggr)
\, . \label{rendf}
\end{equation}
In the limit $\alpha\to\infty$, the right-hand side of (\ref{rendf})
leads to the so-called min-entropy equal to the minus logarithm of
the maximal probability. For $\alpha=1$, both the entropies
(\ref{tsed}) and (\ref{rendf}) reduce to the Shannon entropy,
\begin{equation}
H_{1}(P)=-\sum_{j=1}^{N}p_{j}\ln{p}_{j}
\, . \label{hsed}
\end{equation}
Due to the obvious connection
\begin{equation}
R_{\alpha}(P)=\frac{1}{1-\alpha}\,\ln\bigl\{1+(1-\alpha)H_{\alpha}(P)\bigr\}
\, , \label{reedt}
\end{equation}
each of Tsallis-entropy inequalities may lead to a R\'{e}nyi-entropy
counterpart. In more detail, applications of the Tsallis and
R\'{e}nyi entropies in quantum physics are described in
\cite{bengtsson}.

For the given index of coincidence and $\alpha\in(0,2]$, we have
\begin{equation}
H_{\alpha}(P)\geq\ln_{\alpha}\biggl(\frac{1}{I(P)}\biggr)
 . \label{wwb1}
\end{equation}
This fact follows from Jensen's inequality. The paper \cite{rastrid}
showed the way to improve (\ref{wwb1}) due to information
diagrams. To each probability distribution, we assign the index of
coincidence and the entropy of interest. Using the former as the
abscissa and the latter as the ordinate, each probability
distribution is shown by a point in the plane \cite{harr2001}. The
problem consists in describing the boundaries of the resulting set
of points. In effect, the answer depends on the actual number of
nonzero probabilities. It is known that the lower-bounding curve
passes through points of the form
$\bigl(k^{-1},\ln_{\alpha}(k)\bigr)$ with integer $k\geq1$. It
corresponds to the uniform probability distribution with $k$
elements.

It was demonstrated for information diagrams with the Shannon
entropy that the theoretically best lower-bounding curve is not
smooth \cite{harr2001}. Naturally, the same feature takes place for
the Tsallis $\alpha$-entropy. It can be shown that a smoothness is
violated in the points with abscissas $X=k^{-1}$. In more detail,
this question is explained in \cite{rastrid,raplq}. Actually, the
inequality (\ref{wwb1}) is improved by replacing the graph of smooth
function $X\mapsto\ln_{\alpha}\bigl(X^{-1}\bigr)$ with the polygonal
line connecting the points with abscissas $X=k^{-1}$ and ordinates
$\ln_{\alpha}(k)$ for integer $k\geq1$. New graph shows the
piecewise linear function $X\mapsto{L}_{\alpha}(X)$ expressed as
\cite{rastrid}
\begin{equation}
L_{\alpha}(X)=u_{\alpha{k}}-v_{\alpha{k}}\,X
\, , \qquad
X\in\biggl[\frac{1}{k+1}\, ,\frac{1}{k}\biggr]
\, . \label{elbk}
\end{equation}
Here, the coefficients $u_{\alpha{k}}$ and $v_{\alpha{k}}$ read as
\begin{align}
u_{\alpha{k}}&=(k+1)\ln_{\alpha}(k+1)-k\ln_{\alpha}(k)
\, , \label{abak1}\\
v_{\alpha{k}}&=k(k+1)\bigl\{\ln_{\alpha}(k+1)-\ln_{\alpha}(k)\bigr\}
\, . \label{abak2}
\end{align}
Due to the form of the polygonal line, the function
$X\mapsto{L}_{\alpha}(X)$ is decreasing and convex. It then holds
for $\alpha\in(0,2]$ that \cite{rastrid}
\begin{equation}
H_{\alpha}(P)\geq{L}_{\alpha}\bigl(I(P)\bigr)
\, . \label{wb2}
\end{equation}
This inequality is not tight, except for points of the form
$\bigl(k^{-1},\ln_{\alpha}(k)\bigr)$. For $\alpha=1$, the inequality
(\ref{wb2}) reduces to one of the main results of \cite{harr2001}.
We will apply (\ref{wb2}) to derive uncertainty relations for
quantum measurements of interest including conical $2$-designs.

Information diagrams have also been applied to the maximal
probability. Each diagram now fills so that the ordinate shows
values of the maximal probability \cite{rastrid}. The upper bound on
the maximal probability at the given index of coincidence was
presented in \cite{rastmubs}. Let $X\mapsto\varLambda_{p}(X)$ be a
piecewise smooth function defined as
\begin{equation}
\varLambda_{p}(X)=\frac{1}{k}\,\biggl(1+\sqrt{\frac{kX-1}{k-1}}\>\biggr)
\, , \qquad
X\in\biggl[\frac{1}{k}\, ,\frac{1}{k-1}\biggr]
\, . \label{gpdf}
\end{equation}
This function describes the boundary of the corresponding
information diagram from below \cite{rastrid}. Hence, we obtain a
two-sided estimate of the maximal probability, viz.
\begin{equation}
\varLambda_{p}\bigl(I(P)\bigr)\leq
\underset{1\leq{j}\leq{N}}{\max}\,p_{j}\leq
\frac{1}{N}\,\Bigl(1+\sqrt{N-1}\,\sqrt{NI(P)-1}\,\Bigr)
\, , \label{pat0}
\end{equation}
where $N$ is the number of possible outcomes. The left-hand side of
(\ref{pat0}) is stronger than an obvious estimate
$I(P)\leq\max\,p_{j}$. The latter is tight only in points of the
form $\bigl(k^{-1},k^{-1}\bigr)$ with integer $k\geq1$. Each of
these points of the diagram corresponds to a uniform probability
distribution \cite{rastrid,raplq}.

One of the results of \cite{raplq} allows us to estimate sums of
two probabilities from above at the given index of coincidence. Let
the index of coincidence of the distribution
$P=(p_{1},\ldots,p_{N})$ obey $N^{-1}\leq{I}(P)\leq2^{-1}$. It then
holds that
\begin{equation}
\underset{i\neq{j}}{\max}\,\bigl\{p_{i}+p_{j}\bigr\}
\leq\frac{1}{N}\,
\Bigl(2+\sqrt{2N-4}\,\sqrt{NI(P)-1}\,\Bigr)
\, . \label{in12}
\end{equation}
The inequality (\ref{in12}) is formally correct for all acceptable
values of $I(P)$, but a nontrivial bound is obtained only for
$I(P)<2^{-1}$. This inequality takes place for sufficiently large
number of outcomes. For example, it holds for a complete set of
$d+1$ MUBs in dimension $d\geq3$ \cite{raplq}.

\section{Main results}\label{sec3}

This section presents uncertainty relations for quantum measurements
from generalized equiangular tight frames including conical $2$-designs.
One shall address the entropic formulation as well as the case of
probabilities of separate outcomes. All the derived relations will
be exemplified.

\subsection{Entropic uncertainty relations}\label{sbc31}

Informational entropies \cite{deutsch,maass88} provide a flexible
way to pose the uncertainty principle for finite-dimensional
systems. In some respects, the entropic formulation have advantages
in comparison with the traditional one \cite{robertson}. Entropic
uncertainty relations are important not only from the conceptual
viewpoint, but also for a lot of applications \cite{ww10,cbtw17}. In
particular, such relations are used in derivation of criteria to
detect entanglement \cite{olew,olew1} and steerability
\cite{brun18,cug18,cugen}. Statistics of the measurements of
interest can be interpreted in two ways. First, it deals with a
single generalized equiangular measurement. Second, one links to a
set of generalized symmetric measurements. For these measurements
\cite{getf}, the involved sums of squared probabilities satisfy a
certain relation. This fact leads to a lot of consequences. To avoid
bulky expressions, we will use the positive numbers
\begin{align}
\omega_{\mu}&=\frac{\con^{-1}}{x_{\mu}-y_{\mu}}=\frac{\con^{-1}\gamma_{\mu}^{2}}{a_{\mu}^{2}(b_{\mu}-c_{\mu})}
\ , \label{ommu}\\
\con&=\sum_{\mu=1}^{M}\frac{1}{x_{\mu}-y_{\mu}}
\ . \label{omcu}
\end{align}
By construction, one has
\begin{equation}
\sum_{\mu=1}^{M}\omega_{\mu}=1
\, . \label{omsu}
\end{equation}
Thus, the number (\ref{ommu}) can be treated as a weight related to
the $\mu$-th equiangular line. The following statement takes place.

\newtheorem{prn1}{Proposition}
\begin{prn1}\label{thm1}
Let $\en=\bigl\{\cle_{\mu}\bigr\}$ be a set of $M$ generalized
symmetric POVMs assigned to a generalized equiangular measurement
due to (\ref{linr}). For $\alpha\in(0,2]$, we have
\begin{equation}
\sum_{\mu=1}^{M}\omega_{\mu}\,H_{\alpha}(\cle_{\mu};\bro)\geq{L}_{\alpha}
\biggl(
\frac{d\!\;\tr(\bro^{2})-1}{\con{d}}
+\frac{1}{d}\,\sum_{\mu=1}^{M}w_{\mu}\omega_{\mu}
\biggr)
\, . \label{ep1}
\end{equation}
Let $\clp=\bigcup\,\clp_{\mu}$ be a generalized equiangular
measurement whose elements form a conical $2$-design with the
parameters (\ref{kaps}). For $\alpha\in(0,2]$, it holds that
\begin{equation}
H_{\alpha}(\clp;\bro)\geq{L}_{\alpha}\bigl(S\!\;\tr(\bro^{2})+(\sigma-S)d^{-1}\bigr)
\, . \label{ep2}
\end{equation}
\end{prn1}

{\bf Proof.} To prove (\ref{ep1}), we denote (\ref{emt2}) by
$I_{\mu}$ and further write
\begin{align}
\sum_{\mu=1}^{M}\omega_{\mu}\,H_{\alpha}(\cle_{\mu};\bro)
&\geq\sum_{\mu=1}^{M}\omega_{\mu}\,L_{\alpha}(I_{\mu})
\label{stp1}\\
&\geq{L}_{\alpha}
\biggl(\,\sum_{\mu=1}^{M}\omega_{\mu}\,I_{\mu}\biggr)
\, . \label{stp2}
\end{align}
Here, the step (\ref{stp1}) follows from (\ref{wb2}) and the step
(\ref{stp2}) follows from convexity of the function
$X\mapsto{L}_{\alpha}(X)$. Combining (\ref{emt1}) with
(\ref{stp2}) completes the proof of (\ref{ep1}), since the function
$X\mapsto{L}_{\alpha}(X)$ decreases.

For a conical $2$-design from collection of $M$ generalized
equiangular lines, the author of \cite{getf} proved that
\begin{equation}
\sum_{\mu=1}^{M}\sum_{j=1}^{N_{\mu}}
\tr(\pq_{\mu,j}\bro)^{2}=S\!\;\tr(\bro^{2})+\frac{\sigma-S}{d}
\ . \label{summj}
\end{equation}
Substituting (\ref{summj}) into (\ref{wb2}) immediately gives (\ref{ep2}).
$\blacksquare$

The statement of Proposition \ref{thm1} gives Tsallis-entropy
uncertainty relations for measurements assigned to a generalized
equiangular measurement. In particular, the inequality (\ref{ep1})
is an entropic uncertainty relation with average Tsallis
$\alpha$-entropy. A distinction from the previous considerations is
that averaging in the left-hand side of (\ref{ep1}) is taken with
generally unequal weights. This is a flip side of possibility to
manipulate the coefficients of a generalized equiangular
measurement. In contrast to (\ref{ep1}), the inequality (\ref{ep2})
deals with a single measurement. Instead of (\ref{emt1}), the
corresponding index of coincidence is calculated exactly due to
(\ref{summj}). The weights $\omega_{\mu}$ in the left-hand side of
(\ref{ep1}) are determined by the coefficients of a generalized
equiangular measurement. Since $\tr(\bro^{2})=1$ for a pure state,
the inequalities (\ref{ep1}) and (\ref{ep2}) then give
\begin{align}
\sum_{\mu=1}^{M}\omega_{\mu}\,H_{\alpha}(\cle_{\mu};\bro)&\geq{L}_{\alpha}
\biggl(
\frac{d-1}{\con{d}}
+\frac{1}{d}\,\sum_{\mu=1}^{M}w_{\mu}\omega_{\mu}
\biggr)
\, , \label{ep1p}\\
H_{\alpha}(\clp;\bro)&\geq{L}_{\alpha}\bigl(S+(\sigma-S)d^{-1}\bigr)
\, . \label{ep2p}
\end{align}
These relations actually hold for all states, because the function
$X\mapsto{L}_{\alpha}(X)$ decreases. Such inequalities are typically
used in criteria to detect nonclassical correlations. Together with
the weights $\omega_{\mu}$, the right-hand side of (\ref{ep1p}) can
be treated as an estimation of Tsallis entropies related to $M$
generalized symmetric POVMs. In a similar vein, the right-hand side
of (\ref{ep2p}) is an entropic characteristic of the given conical
$2$-design. The next formulation of uncertainty relations uses the
R\'{e}nyi entropies.

\newtheorem{prn2}[prn1]{Proposition}
\begin{prn2}\label{thm2}
Let $\en=\bigl\{\cle_{\mu}\bigr\}$ be a set of $M$ generalized
symmetric POVMs assigned to a generalized equiangular measurement
due to (\ref{linr}). For $\alpha\in[1,2]$, we have
\begin{equation}
\sum_{\mu=1}^{M}\omega_{\mu}\,R_{\alpha}(\cle_{\mu};\bro)\geq
\frac{1}{1-\alpha}\,\ln\biggl\{1+(1-\alpha)
L_{\alpha}\biggl(
\frac{\tr(\bro^{2})\!\;d-1}{\con{d}}
+\frac{1}{d}\,\sum_{\mu=1}^{M}w_{\mu}\omega_{\mu}
\biggr)\biggr\}
\, . \label{rp1}
\end{equation}
Let $\clp=\bigcup\,\clp_{\mu}$ be a generalized equiangular
measurement whose elements form a conical $2$-design with the
parameters (\ref{kaps}). For $\alpha\in(0,2]$, it holds that
\begin{equation}
R_{\alpha}(\clp;\bro)\geq
\frac{1}{1-\alpha}\,\ln\Bigl\{1+(1-\alpha)L_{\alpha}\bigl(S\!\;\tr(\bro^{2})+(\sigma-S)d^{-1}\bigr)\Bigr\}
\, . \label{rp2}
\end{equation}
\end{prn2}

{\bf Proof.} To prove (\ref{rp1}), one writes
\begin{align}
\sum_{\mu=1}^{M}\omega_{\mu}\,R_{\alpha}(\cle_{\mu};\bro)
&=\frac{1}{1-\alpha}\,\sum_{\mu=1}^{M}
\omega_{\mu}\ln\bigl\{1+(1-\alpha)H_{\alpha}(\cle_{\mu};\bro)\bigr\}
\label{et1r}\\
&\geq
\frac{1}{1-\alpha}\,\sum_{\mu=1}^{M}
\omega_{\mu}\ln\bigl\{1+(1-\alpha)L_{\alpha}\bigl(I_{\mu}\bigr)\bigr\}
\, . \label{et2r}
\end{align}
Here, the step (\ref{et1r}) follows from (\ref{reedt}) and the step
(\ref{et2r}) follows from (\ref{wb2}). The formula (\ref{et2r})
remains valid for $\alpha\in(0,2]$. If the function $X\mapsto{f}(X)$
is convex, and the function $Y\mapsto{g}(Y)$ is increasing and
convex, then the composition $X\mapsto{g}\bigl(f(X)\bigr)$ is convex
as well. Then the features of $X\mapsto{L}_{\alpha}(X)$ and
$Y\mapsto(1-\alpha)^{-1}\ln\bigl\{1+(1-\alpha)Y\bigr\}$ imply
convexity of the function \cite{raplq}
\begin{equation}
X\mapsto\frac{1}{1-\alpha}\,\ln\bigl\{1+(1-\alpha)L_{\alpha}(X)\bigr\}
 \label{decx}
\end{equation}
for $\alpha\in[1,2]$. In this range, the inequalities (\ref{et2r})
and (\ref{emt1}) lead to (\ref{rp1}), since the function
(\ref{decx}) decreases.

When the given generalized equiangular measurement is a conical
$2$-design, the index of coincidence is fixed by (\ref{summj}) that
results in (\ref{ep2}). Combining the latter with (\ref{reedt}) and
increase of the function
$Y\mapsto(1-\alpha)^{-1}\ln\bigl\{1+(1-\alpha)Y\bigr\}$ immediately
leads to (\ref{rp2}). $\blacksquare$

The inequalities (\ref{rp1}) and (\ref{rp2}) are R\'{e}nyi-entropy
uncertainty relations for measurements assigned to a generalized
equiangular measurement. The first formula estimates from below the
R\'{e}nyi entropy averaged over POVMs with the actual weights
(\ref{ommu}). In contrast to (\ref{rp2}), it is proved only for
$\alpha\in[1,2]$. The latter follows from convexity properties of
the functions involved into consideration. The inequality
(\ref{rp2}) deals with a single generalized equiangular measurement.
It holds for $\alpha\in(0,2]$ similarly to (\ref{ep2}). For a pure
state, the inequalities (\ref{rp1}) and (\ref{rp2}) give
\begin{align}
\sum_{\mu=1}^{M}\omega_{\mu}\,R_{\alpha}(\cle_{\mu};\bro)&\geq
\frac{1}{1-\alpha}\,\ln\biggl\{1+(1-\alpha)
L_{\alpha}\biggl(
\frac{d-1}{\con{d}}
+\frac{1}{d}\,\sum_{\mu=1}^{M}w_{\mu}\omega_{\mu}
\biggr)\biggr\}
\, , \label{rp1p}\\
R_{\alpha}(\clp;\bro)&\geq
\frac{1}{1-\alpha}\,\ln\Bigl\{1+(1-\alpha)L_{\alpha}\bigl(S+(\sigma-S)d^{-1}\bigr)\Bigr\}
\, . \label{rp2p}
\end{align}
Similarly to (\ref{ep1p}) and (\ref{ep2p}), these relations hold for
all states and have potential applications. The right-hand sides of
(\ref{rp1p}) and (\ref{rp2p}) can be interpreted as an estimate of
R\'{e}nyi entropies assigned to the corresponding measurements.

\subsection{Fine-grained uncertainty relations}\label{sbc32}

For a pair of observables, uncertainty relations of the
Landau--Pollak type are expressed in terms of the maximal
probabilities. Its quantum-mechanical interpretation was emphasized
in \cite{maass88}, since the original formulation \cite{landau61}
concerned signal analysis. Relations of the Landau--Pollak type can
also be considered as fine-grained uncertainty relations. The
authors of \cite{oppew10} discussed the role of such relations
dealing with a particular combination of the outcomes. This paper
deals with POVMs such that the number of outcomes exceeds
dimensionality. Then a nontrivial bound from above holds already for
a single probability \cite{rastmubs}. Namely, the following
statement takes place.

\newtheorem{prn3}[prn1]{Proposition}
\begin{prn3}\label{thm3}
Let $\en=\bigl\{\cle_{\mu}\bigr\}$ be a set of $M$ generalized
symmetric POVMs assigned to a generalized equiangular measurement
due to (\ref{linr}), and let $N_{\mu}=N$ for all $\mu=1,\ldots,M$.
It holds that
\begin{equation}
\sum_{\mu=1}^{M}\omega_{\mu}\,\underset{1\leq{j}\leq{N}}{\max}\,\tr(\me_{\mu,j}\bro)
\leq
\frac{1}{N}\,\biggl\{1+\sqrt{N-1}\,\biggl(\frac{Nd\!\;\tr(\bro^{2})-N}{\con{d}}
+\frac{N}{d}\,\sum_{\mu=1}^{M}w_{\mu}\omega_{\mu}-1\biggr)^{\!1/2}\,\biggr\}
\, . \label{pat01}
\end{equation}
Let $\clp=\bigcup\,\clp_{\mu}$ be a generalized equiangular
measurement whose elements form a conical $2$-design with the
parameters (\ref{kaps}). The maximal probability satisfies
\begin{equation}
\varLambda_{p}\bigl(S\!\;\tr(\bro^{2})+(\sigma-S)d^{-1}\bigr)\leq
\underset{\mu,j}{\max}\,\tr(\pq_{\mu,j}\bro)
\leq
\frac{1}{K}\,\biggl\{1+\sqrt{K-1}\,\Bigl(KS\!\;\tr(\bro^{2})+K(\sigma-S)d^{-1}-1\Bigr)^{\!1/2}\,\biggr\}
\, , \label{pat02}
\end{equation}
where the total number of outcomes
\begin{equation}
K=\sum_{\mu=1}^{M}N_{\mu}
\, . \label{ntot}
\end{equation}
\end{prn3}

{\bf Proof.} The right-hand side of (\ref{pat0}) is a concave
function of $I(P)$. For a set of $M$ POVMs with the same number of
outcomes, we can write
\begin{equation}
\sum_{\mu=1}^{M}\omega_{\mu}\,\underset{1\leq{j}\leq{N}}{\max}\,\tr(\me_{\mu,j}\bro)
\leq\sum_{\mu=1}^{M}
\frac{\omega_{\mu}}{N}\,\Bigl(1+\sqrt{N-1}\,\sqrt{NI_{\mu}-1}\,\Bigr)
\leq
\frac{1}{N}\,\biggl\{1+\sqrt{N-1}\,\biggl(N\sum_{\mu=1}^{M}\omega_{\mu}I_{\mu}-1\biggr)^{\!1/2}\,\biggr\}
\, . \nonumber
\end{equation}
Combining the latter with (\ref{emt1}) completes the proof of
(\ref{pat01}). The complementarity relation (\ref{pat02}) is
obtained by substituting (\ref{summj}) into the two-sided estimate
(\ref{pat0}). $\blacksquare$

The statement of Proposition \ref{thm3} imposes restrictions on
maximal probabilities of the considered measurement. The result
(\ref{pat01}) estimates from above the maximal probability averaged
over POVMs with the weights (\ref{ommu}). In general, these weight
are not equal. The second result (\ref{pat02}) provides estimations
of the maximal probability from below and above simultaneously.
These results could be converted into inequalities with the
corresponding min-entropies. We refrain from presenting the details
here. The state-independent counterparts of (\ref{pat01}) and
(\ref{pat02}) respectively read as
\begin{align}
\sum_{\mu=1}^{M}\omega_{\mu}\,\underset{1\leq{j}\leq{N}}{\max}\,\tr(\me_{\mu,j}\bro)
&\leq
\frac{1}{N}\,\biggl\{1+\sqrt{N-1}\,\biggl(\frac{Nd-N}{\con{d}}
+\frac{N}{d}\,\sum_{\mu=1}^{M}w_{\mu}\omega_{\mu}-1\biggr)^{\!1/2}\,\biggr\}
\, , \label{pat01p}\\
\underset{\mu,j}{\max}\,\tr(\pq_{\mu,j}\bro)
&\leq
\frac{1}{K}\,\biggl\{1+\sqrt{K-1}\,\Bigl(KS+K(\sigma-S)d^{-1}-1\Bigr)^{\!1/2}\,\biggr\}
\, . \label{pat02p}
\end{align}
These inequalities are obtained by replacing $\tr(\bro^{2})$ with
its maximal value one. Note that only an estimate from above appears
in (\ref{pat02p}). The formulas (\ref{pat01p}) and (\ref{pat02p})
allow us to estimate the maximal probabilities from above in terms
of the characteristics of POVMs solely.

To formulate relations for the maximal sum of two probabilities, we
introduce auxiliary function
\begin{equation}
F_{N}(X)=
\begin{cases}
 N^{-1}\bigl(2+\sqrt{2N-4}\,\sqrt{NX-1}\,\bigr) \, , & \text{if}\ N^{-1}\leq{X}\leq2^{-1}\, , \\
 1 \, , & \text{if}\ 2^{-1}\leq{X}\leq1\, .
\end{cases}
\label{fnx}
\end{equation}
One shall now apply (\ref{in12}) to measurements assigned to a
generalized equiangular measurement.

\newtheorem{prn4}[prn1]{Proposition}
\begin{prn4}\label{thm4}
Let $\en=\bigl\{\cle_{\mu}\bigr\}$ be a set of $M$ generalized
symmetric POVMs assigned to a generalized equiangular measurement
due to (\ref{linr}), and let $N_{\mu}=N$ for all $\mu=1,\ldots,M$.
It holds that
\begin{equation}
\sum_{\mu=1}^{M}\omega_{\mu}\,\underset{i\neq{j}}{\max}\,\bigl\{\tr(\me_{\mu,i}\bro)+\tr(\me_{\mu,j}\bro)\bigr\}
\leq{F}_{N}\biggl(\frac{d\!\;\tr(\bro^{2})-1}{\con{d}}
+\frac{1}{d}\,\sum_{\mu=1}^{M}w_{\mu}\omega_{\mu}\biggr)
\, . \label{int1}
\end{equation}
Let $\clp=\bigcup\,\clp_{\mu}$ be a generalized equiangular
measurement whose elements form a conical $2$-design with the
parameters (\ref{kaps}); it holds that
\begin{equation}
\underset{(\mu,i)\neq(\nu,j)}{\max}\,\bigl\{\tr(\pq_{\mu,i}\bro)+\tr(\pq_{\nu,j}\bro)\bigr\}
\leq{F}_{K}\!\bigl(S\!\;\tr(\bro^{2})+(\sigma-S)d^{-1}\bigr)
\, , \label{int2}
\end{equation}
where the total number of outcomes $K$ is defined by (\ref{ntot}).
\end{prn4}

{\bf Proof.} By construction, the function $X\mapsto{F}_{N}(X)$ is
increasing and concave. Due to (\ref{in12}), we therefore have
\begin{equation}
\sum_{\mu=1}^{M}\omega_{\mu}\,
\underset{i\neq{j}}{\max}\,\bigl\{\tr(\me_{\mu,i}\bro)+\tr(\me_{\mu,j}\bro)\bigr\}
\leq\sum_{\mu=1}^{M}
\omega_{\mu}\,F_{N}(I_{\mu})
\leq
{F}_{N}\biggl(\,\sum_{\mu=1}^{M}\omega_{\mu}\,I_{\mu}\biggr)
\, . \nonumber
\end{equation}
Combining the latter with (\ref{emt1}) completes the proof of
(\ref{int1}). The result (\ref{int2}) directly follows from
(\ref{in12}) and (\ref{summj}). $\blacksquare$

The statement of Proposition \ref{thm4} is an example of
fine-grained uncertainty relations with the use of probabilities of
two separate outcomes. It is not a surprise that the inequality
(\ref{int1}) realizes averaging with the weights (\ref{ommu})
generally unequal. In the paper \cite{raplq}, similar relations with
equal weights were formulated for measurements assigned to a
projective $2$-design. In particular, the case of $d+1$ MUBs was
addressed assuming that $d$ is a prime power. Thus, Proposition
\ref{thm4} provides an extension of recent results to generalized
equiangular measurements. To convert (\ref{int1}) and (\ref{int2})
into the state-independent form, we replace $\tr(\bro^{2})$ with its
maximal value one, whence
\begin{align}
\sum_{\mu=1}^{M}\omega_{\mu}\,\underset{i\neq{j}}{\max}\,\bigl\{\tr(\me_{\mu,i}\bro)+\tr(\me_{\mu,j}\bro)\bigr\}
\leq{F}_{N}\biggl(\frac{d-1}{\con{d}}
+\frac{1}{d}\,\sum_{\mu=1}^{M}w_{\mu}\omega_{\mu}\biggr)
\, , \label{int1p}\\
\underset{(\mu,i)\neq(\nu,j)}{\max}\,\bigl\{\tr(\pq_{\mu,i}\bro)+\tr(\pq_{\nu,j}\bro)\bigr\}
\leq{F}_{K}\!\bigl(S+(\sigma-S)d^{-1}\bigr)
\, . \label{int2p}
\end{align}
It would be interesting to examine applications of these
inequalities in quantum information, including schemes to detect
entanglement and steerability.

\subsection{Examples}\label{sbc33}

Let us illustrate the derived relations with some examples. It is
instructive to begin with a qubit. Its density matrix reads as
\begin{equation}
\bro=\frac{1}{2}\,\bigl(\pen_{2}+r_{x}\bau_{x}+r_{y}\bau_{y}+r_{z}\bau_{z}\bigr)
\, , \label{dnmatr}
\end{equation}
where $r_{x}=\sin\theta\cos\varphi$, $r_{y}=\sin\theta\sin\varphi$
and $r_{z}=\cos\theta$ are the components of the Bloch vector. By
$\theta$ and $\varphi$, we mean usual angles on the Bloch sphere.
Three eigenbases of the Pauli matrices $\bau_{x}$, $\bau_{y}$ and
$\bau_{z}$ are mutually unbiased. This case deals with three von
Neumann measurements. Here, the weights (\ref{ommu}) are equal to
one third. The six states of the three eigenbases also form a
$3$-design \cite{guhne20} and, herewith, a $2$-design. The resulting
POVM consists of elements of the form
$3^{-1}|b_{\mu,j}\rangle\langle{b}_{\mu,j}|$ with $\mu=x,y,z$ and
$j=\pm$. It follows from (\ref{ep1}) and (\ref{ep2}) that, for
$\alpha\in[0,2]$,
\begin{align}
\frac{1}{3}\sum_{\mu=x,y,z}H_{\alpha}(\clb_{\mu};\bro)\geq{L}_{\alpha}
\biggl(
\frac{3+r^{2}}{6}
\biggr)
\, , \label{eqb1}\\
H_{\alpha}(\clp;\bro)\geq{L}_{\alpha}
\biggl(
\frac{3+r^{2}}{18}
\biggr)
\, , \label{eqb2}
\end{align}
where $r^{2}=r_{x}^{2}+r_{y}^{2}+r_{z}^{2}$ is square of the Bloch
vector. The uncertainty relations (\ref{eqb1}) and  (\ref{eqb2})
were actually reported in \cite{raplq}. A utility of these
inequalities was also discussed therein.

The next example deals with unequal weights (\ref{ommu}). Following
\cite{getf}, we consider the operators
\begin{align}
\me_{1,1}&=
\begin{pmatrix}
 1 & 0 \\
 0 & 0
\end{pmatrix}
 , \qquad
\me_{1,2}=
\begin{pmatrix}
 0 & 0 \\
 0 & 1
\end{pmatrix}
 , \label{mu1}\\
\me_{2,1}&=\frac{1}{3}
\begin{pmatrix}
 1 & -\!\;\iu \\
 \iu & 1
\end{pmatrix}
 , \qquad
\me_{2,2}=\frac{1}{6}
\begin{pmatrix}
 2 & \sqrt{3}+\iu \\
 \sqrt{3}-\iu & 2
\end{pmatrix}
 , \qquad
\me_{2,3}=\frac{1}{6}
\begin{pmatrix}
 2 & -\!\;\sqrt{3}+\iu \\
 -\!\;\sqrt{3}-\iu & 2
\end{pmatrix}
 . \label{mu2}
\end{align}
It can be checked that the parameters (\ref{sm1})--(\ref{sm4}) read as
\begin{equation}
w_{1}=1
\, , \qquad
w_{2}=\frac{2}{3}
\, , \qquad
x_{1}=1
\, , \qquad
x_{2}=\frac{4}{9}
\ , \qquad
y_{1}=0
\ , \qquad
y_{2}=\frac{1}{9}
\ , \qquad
z_{12}=\frac{1}{3}
\, . \nonumber
\end{equation}
Let us define the five elements of generalized equiangular measurement as
\begin{equation}
\pq_{1,i}=\frac{2}{5}\,\me_{1,i}
\, , \qquad
\pq_{2,j}=\frac{3}{5}\,\me_{2,j}
\, . \label{pqem}
\end{equation}
We further calculate
\begin{equation}
a_{1}=a_{2}=\frac{2}{5}
\, , \qquad
\tr(\pq_{\mu,j}^{2})=\frac{4}{25}
\, , \qquad
b_{1}=b_{2}=1
\, , \qquad
c_{1}=0
\, , \qquad
c_{2}=\frac{1}{4}
\ , \qquad
\omega_{1}=\frac{1}{4}
\ , \qquad
\omega_{2}=\frac{3}{4}
\ , \nonumber
\end{equation}
and $\con=4$. These values should be kept in mind when looking at the graphs.

\begin{figure}
\includegraphics[height=7.2cm]{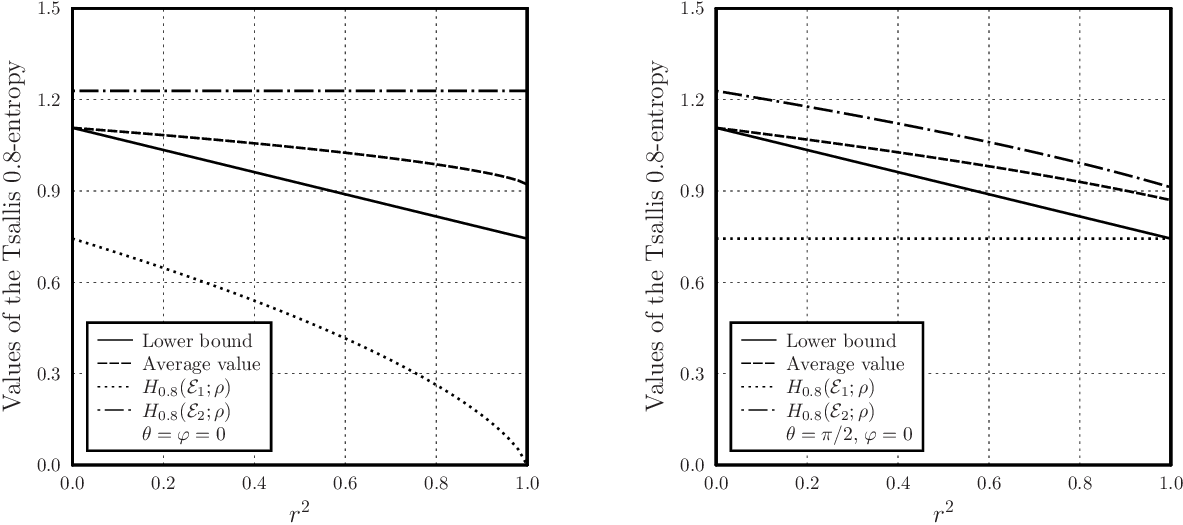}
\caption{\label{fig1} Values of the Tsallis $0.8$-entropy versus square of the Bloch vector, whose direction coincides with the $z$-axis on the left and the $x$-axis on the right.}
\end{figure}

It is instructive to compare the entropies of interest for two POVMs
$\bigl\{\me_{1,1},\me_{1,2}\bigr\}$ and
$\bigl\{\me_{2,1},\me_{2,2},\me_{2,3}\bigr\}$ separately, the
average one and the lower bound (\ref{ep1}). These entropic values
are shown in Fig. \ref{fig1} for $\alpha=0.8$ and the two directions
of the Bloch vector. The abscissa shows square of the Bloch vector.
Of course, the curve of the right-hand side of (\ref{ep1}) is the
same in both the left and right boxes. It is seen that the lower
bound is saturated for the maximally mixed state. The actual
direction of the Bloch vector significantly governs the two
entropies $H_{\alpha}(\cle_{1};\bro)$ and
$H_{\alpha}(\cle_{2};\bro)$. At the same time, the average entropic
value is affected not so essentially. The derived lower bound seems
to be sufficiently tight. The deviation is maximal for pure states.
On a relative scale, it is around 19.3 \% on the left and 14.5 \% on
the right. The former is particularly conditioned by the fact that
$H_{\alpha}(\cle_{1};\bro)$ vanishes. With growth of mixedness, the
deviation reduces. This example is interesting due to a combination
of the two POVMs with different numbers of outcomes. Despite of this
choice, the inequality (\ref{ep1}) is not far from optimality.

\begin{figure}
\includegraphics[height=7.2cm]{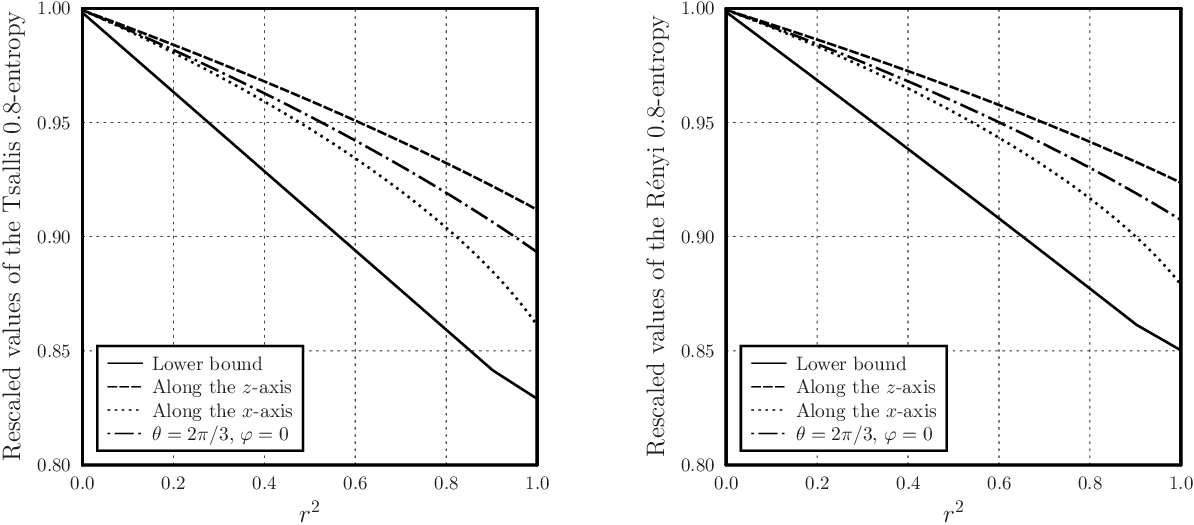}
\caption{\label{fig2} Rescaled values of the $0.8$-entropies versus square of the Bloch vector for three directions: the Tsallis $0.8$-entropy on the left and the R\'{e}nyi $0.8$-entropy on the right.}
\end{figure}

To exemplify the case of conical $2$-design, we use the operators
\cite{getf}
\begin{align}
\pq_{1,1}=\frac{3-\sqrt{5}}{8}
\begin{pmatrix}
 \sqrt{5}-\sqrt{3} & 0 \\
 0 & \sqrt{5}+\sqrt{3}
\end{pmatrix}
 , \qquad
\pq_{1,2}=\frac{3-\sqrt{5}}{8}
\begin{pmatrix}
 \sqrt{5}+\sqrt{3} & 0 \\
 0 & \sqrt{5}-\sqrt{3}
\end{pmatrix}
 , \label{pmu1}\\
\pq_{2,1}=\frac{3-\sqrt{5}}{8}
\begin{pmatrix}
 2 & q \\
 q^{*} & 2
\end{pmatrix}
 , \qquad
\pq_{2,2}=\frac{3-\sqrt{5}}{8}
\begin{pmatrix}
 2 & -\!\;\iu{q}^{*} \\
 \iu{q} & 2
\end{pmatrix}
 , \qquad
\pq_{2,3}=\frac{3-\sqrt{5}}{4\sqrt{2}}
\begin{pmatrix}
 \sqrt{2} & 1-\iu \\
 1+\iu & \sqrt{2}
\end{pmatrix}
 . \label{pmu2}
\end{align}
where $-\!\;q=\bigl(2+\sqrt{3}+\iu\bigr)\sqrt{2-\sqrt{3}}$. The
required parameters read as
\begin{align}
\gamma_{1}&=\frac{3\sqrt{5}-5}{4}
\ , \qquad
\gamma_{2}=1-\gamma_{1}
\, , \label{gam12}\\
\tr(\pq_{\mu,j}^{2})&=a_{\mu}^{2}b_{\mu}
=\frac{7-3\sqrt{5}}{2}
\ , \label{trp2}\\
\tr(\pq_{\mu,i}\pq_{\mu,j})&=a_{\mu}^{2}c_{\mu}=\frac{7-3\sqrt{5}}{8}
\qquad (i\neq{j})
\, , \label{trp3}\\
\tr(\pq_{\mu,i}\pq_{\nu,j})&=\frac{7\sqrt{5}-15}{8}
\qquad (\mu\neq\nu)
\, . \label{trp4}
\end{align}
It is interesting that the traces (\ref{trp2})--(\ref{trp4}) are
independent of $\mu$ \cite{getf}. Therefore, we obtain here a
conical $2$-design such that
\begin{equation}
S=\frac{21-9\sqrt{5}}{8}
\ , \qquad
\sigma=\frac{11\bigl(3-\sqrt{5}\bigr)^{2}}{16}
\ . \label{ssig}
\end{equation}
Dealing with a single measurement, one can rescale each of the used
entropies by its maximal possible value. Then the curves all reduce
to the same interval. This example includes five outcomes, so that
we divide $H_{\alpha}(\clp;\bro)$ by $\ln_{\alpha}(5)$ and
$R_{\alpha}(\clp;\bro)$ by $\ln5$. The rescaled values of the
$\alpha$-entropies versus square of the Bloch vector are shown in
Fig. \ref{fig2} for $\alpha=0.8$ and the three directions. The lower
bounds are calculated due to (\ref{ep2}) and (\ref{rp2}),
respectively. Similarly to the picture of Fig. \ref{fig1}, the
deviation from the lower bounds is maximal for pure states. For the
Bloch vector along the $z$-axis, it is around 9.1 \% in the Tsallis
case and 7.9 \% in the R\'{e}nyi case. In a relative scale,
deviations are less than for the curves in Fig. \ref{fig1}. Indeed,
this example deals with the number of outcomes that is larger.

To illustrate some of the results for separate probabilities, we
also recall the concept of mutually unbiased measurements (MUMs)
\cite{kalev2014}. Each of them contains $N_{\mu}=d$ elements. The
set $\qn=\{\clq_{1},\ldots,\clq_{M}\}$ is a set of $M$ MUMs of the
efficiency $\varkappa$ in dimension $d$, when
\begin{equation}
\tr(\qp_{\mu,j})=1
\, , \qquad
\tr(\qp_{\mu,i}\qp_{\mu,j})=\delta_{ij}\,\varkappa
+(1-\delta_{ij})\>\frac{1-\varkappa}{d-1}
\ , \qquad
\tr(\qp_{\mu,i}\qp_{\nu,j})=\frac{1}{d}
\qquad (\mu\neq\nu)
\, . \label{mum13}
\end{equation}
The authors of \cite{kalev2014} showed how to build a complete set
of $d+1$ mutually unbiased measurements in $\hh_{d}$. In general,
the parameter $\varkappa$ obeys $d^{-1}\leq\varkappa\leq1$. The
maximal efficiency $\varkappa=1$ corresponds to mutually unbiased
bases (MUBs). In the considered example, the
parameters (\ref{sm1})--(\ref{sm4}) read as
\begin{equation}
w_{\mu}=1
\, ,    \qquad
x_{\mu}=\varkappa
\, ,    \qquad
y_{\mu}=\frac{1-\varkappa}{d-1}
\ ,     \qquad
z_{\mu\nu}=\frac{1}{d}
\ .     \label{mupar}
\end{equation}
It was proved in \cite{rastopn} that
\begin{equation}
\frac{1}{M}\,\sum_{\mu=1}^{M}I(\clq_{\mu};\bro)\leq\frac{M-1}{Md}+\frac{1-\varkappa+(\varkappa{d}-1)\!\;\tr(\bro^{2})}{M(d-1)}
\ . \nonumber
\end{equation}
For $M=d+1$, this inequality is saturated so that \cite{rastopn}
\begin{equation}
\frac{1}{d+1}\,\sum_{\mu=1}^{d+1}I(\clq_{\mu};\bro)=\frac{1}{d+1}+\frac{1-\varkappa+(\varkappa{d}-1)\!\;\tr(\bro^{2})}{d^{2}-1}
\ . \label{ubp2}
\end{equation}
For a set of $d+1$ MUMs, we obviously have $\omega_{\mu}=(d+1)^{-1}$, so
that the inequality (\ref{pat01}) gives
\begin{equation}
\frac{1}{d+1}\,\sum_{\mu=1}^{d+1}\underset{1\leq{j}\leq{d}}{\max}\,\tr(\qp_{\mu,j}\bro)
\leq\frac{1}{d}+\frac{1}{d}\,\sqrt{
\frac{(\varkappa{d}-1)\bigl[d\!\;\tr(\bro^{2})-1\bigr]}{d+1}}
\ . \label{mumap}
\end{equation}
Further, the inequality (\ref{int1}) leads to
\begin{equation}
\frac{1}{d+1}\,
\sum_{\mu=1}^{d+1}\underset{i\neq{j}}{\max}\,\bigl\{\tr(\qp_{\mu,i}\bro)+\tr(\qp_{\mu,j}\bro)\bigr\}
\leq\frac{2}{d}+\frac{\sqrt{2d-4}}{d}\,\sqrt{
\frac{(\varkappa{d}-1)\bigl[d\!\;\tr(\bro^{2})-1\bigr]}{d^{2}-1}}
\ . \label{mumbp}
\end{equation}
For the maximal efficiency $\varkappa=1$ corresponding to MUBs the
inequality (\ref{mumbp}) reduces to one of the results proved in
\cite{raplq}. Then its right-hand side is strictly less than 1 for
$d\geq3$ and a state that is not pure. For $\varkappa<1$ and $d\geq3$, the
inequality (\ref{mumbp}) is nontrivial for all states. Using MUMs, a
generalized equiangular measurement can be built due to
(\ref{linr}). We refrain from presenting the details here.

\section{Conclusions}\label{sec4}

We have derived uncertainty relations for measurements assigned to a
generalized equiangular measurement. Each of such POVMs is built from
generalized equiangular tight frames that are complementary to one
another. This concept recently proposed in \cite{getf} allows us to
manipulate with the involved coefficients in several ways. For
instance, there are generalized equiangular measurement for which
three out of four defining parameters becomes independent of the
measurement operator indices. The role of measurements with a
special inner structure is well known in quantum information theory.
Complete sets of MUBs and SIC-POVMs are probably most known in
this regard. At the same time, the question of existence for such
measurements seems to be very hard. To mitigate these questions, one
can weak some of the imposed conditions. Mutually unbiased
measurements \cite{kalev2014} and general SIC-POVMs \cite{kalev2014}
are examples of such a kind. The paper \cite{getf} proposes further
development in this direction.

The presented results follow from an estimation of the index of
coincidence for POVMs of interest. In addition, several properties
of information characteristics were used to derive uncertainty
relations. The used approach is based on inequalities obtained by
means of information diagrams. In more detail, these questions are
considered in \cite{rastrid,raplq}. Namely, generalized entropies of
both the R\'{e}nyi and Tsallis types were utilized. Uncertainty
relations with probabilities of separate outcomes have also been
addressed. Dealing with average characteristics involves generally
unequal weights. The latter is a flip side of existing freedom to
manipulate the defining parameters. All the new uncertainty
relations are briefly exemplified. Like MUBs and SICs, quantum
measurements from generalized equiangular tight frames can allow one
to detect entanglement and steerability. Applications of the
presented uncertainty relations for such purposes will be considered
in future investigations.

\appendix

\section{A statement on indices of coincidence}\label{incva}

In principle, the required technique was developed by the author of
\cite{getf}. The key distinction of this appendix is that the
considered sets of operators are assumed to be incomplete in
general. Instead of the equality presented in \cite{getf}, the
following inequality takes place.

\newtheorem{lem1}{Lemma}
\begin{lem1}\label{lm1}
Let $\en=\bigl\{\cle_{\mu}\bigr\}$ be a set of $M$ generalized
symmetric POVMs assigned to a generalized equiangular measurement
due to (\ref{linr}), with the $\mu$-th index of coincidence
\begin{equation}
I(\cle_{\mu};\bro)=\sum_{j=1}^{N_{\mu}}\tr(\me_{\mu,j}\bro)^{2}
\, . \label{emt2}
\end{equation}
It holds that
\begin{equation}
\sum_{\mu=1}^{M}\omega_{\mu}\,I(\cle_{\mu};\bro)\leq
\frac{d\!\;\tr(\bro^{2})-1}{\con{d}}
+\frac{1}{d}\,\sum_{\mu=1}^{M}w_{\mu}\omega_{\mu}
\, , \label{emt1}
\end{equation}
where the numbers $\omega_{\mu}$ and $\con$ are defined by
(\ref{ommu}) and (\ref{omcu}), respectively.
\end{lem1}

{\bf Proof.} Instead of (\ref{emt1}), we aim to prove the equivalent
inequality
\begin{equation}
\sum_{\mu=1}^{M}
\frac{1}{a_{\mu}^{2}(b_{\mu}-c_{\mu})}
\,
\biggl(
\,\sum_{j=1}^{N_{\mu}}\tr(\pq_{\mu,j}\bro)^{2}
-\frac{a_{\mu}\gamma_{\mu}}{d}
\biggr)
\leq
\tr(\bro^{2})-f
\ . \label{emt11}
\end{equation}
Their equivalence can be checked immediately due to (\ref{linr}) and
(\ref{abcf}). Let us represent an arbitrary $\bro$ as
\begin{equation}
\bro=\sum_{\mu=1}^{M}\sum_{j=1}^{N_{\mu}}\tr(\pq_{\mu,j}\bro)\mg_{\mu,j}+\bogm
\, . \label{bgmp}
\end{equation}
For the given elements $\pq_{\mu,j}$, the operators $\mg_{\mu,j}$
read as \cite{getf}
\begin{equation}
\mg_{\mu,j}=\frac{1}{a_{\mu}^{2}(b_{\mu}-c_{\mu})}
\,
\biggl\{\pq_{\mu,j}-
\frac{1}{d}
\biggl(
a_{\mu}-\frac{a_{\mu}^{2}(b_{\mu}-c_{\mu})}{M\gamma_{\mu}}
\biggr)\pen_{d}
\biggr\}
\, . \label{mgdf}
\end{equation}
By $\bogm$, we mean in (\ref{bgmp}) potentially nonzero part of
$\bro$ such that $\tr(\mg_{\mu,j}\bogm)=0$ for all $\mu$ and $j$.
Being depending on $\bro$, this term does not appear in
(\ref{emt11}). To check (\ref{bgmp}), one writes
\begin{equation}
\tr(\pq_{\mu,i}\bro)=\sum_{\nu=1}^{M}\sum_{j=1}^{N_{\nu}}\tr(\pq_{\nu,j}\bro)\,\tr(\pq_{\mu,i}\mg_{\nu,j})
\, . \label{check1}
\end{equation}
It needs to show that the right-hand side of this formula indeed
reduces to its left-hand side. Doing some calculations finally
results in the formulas
\begin{align}
\sum_{\nu\neq\mu}\sum_{j=1}^{N_{\nu}}\tr(\pq_{\nu,j}\bro)\,\tr(\pq_{\mu,i}\mg_{\nu,j})
&=\frac{(M-1)a_{\mu}}{Md}
\ , \label{nnm}\\
\sum_{j=1}^{N_{\mu}}\tr(\pq_{\mu,j}\bro)\,\tr(\pq_{\mu,i}\mg_{\mu,j})
&=\tr(\pq_{\mu,i}\bro)-\frac{(M-1)a_{\mu}}{Md}
\ . \label{scn2}
\end{align}
Combining (\ref{nnm}) with (\ref{scn2}) proves (\ref{check1}).

Due to definition of $\bogm$, we see from (\ref{bgmp}) that
\begin{equation}
\tr(\bro^{2})\geq\sum_{\mu=1}^{M}\sum_{\nu=1}^{M}\sum_{i=1}^{N_{\mu}}\sum_{j=1}^{N_{\nu}}\tr(\pq_{\mu,i}\bro)\,\tr(\pq_{\nu,j}\bro)\,\tr\bigl(\mg_{\mu,i}\mg_{\nu,j}\bigr)
\, . \label{trrho2}
\end{equation}
It also follows from (\ref{mgdf}) that
\begin{equation}
\tr\bigl(\mg_{\mu,i}\mg_{\nu,j}\bigr)=\frac{\delta_{\mu\nu}\delta_{ij}a_{\mu}^{2}(b_{\mu}-c_{\mu})+\delta_{\mu\nu}a_{\mu}^{2}(c_{\mu}-f)}{a_{\mu}^{2}(b_{\mu}-c_{\mu})a_{\nu}^{2}(b_{\nu}-c_{\nu})}+\frac{f}{M^{2}\gamma_{\mu}\gamma_{\nu}}
\ . \label{trgg11}
\end{equation}
Substituting (\ref{trgg11}) into the right-hand side of
(\ref{trrho2}) leads to the different sums. By
calculations, the sum with the last term of (\ref{trgg11}) reads as
\begin{equation}
\frac{f}{M^{2}}\sum_{\mu=1}^{M}\,\frac{1}{\gamma_{\mu}}\sum_{i=1}^{N_{\mu}}\tr(\pq_{\mu,i}\bro)\sum_{\nu=1}^{M}\,\frac{1}{\gamma_{\nu}}\sum_{j=1}^{N_{\nu}}\tr(\pq_{\nu,j}\bro)
=\frac{f}{M^{2}}\sum_{\mu=1}^{M}\sum_{\nu=1}^{M}1=f
\, . \label{third}
\end{equation}
Substituting the first fraction in the right-hand side of
(\ref{trgg11}) into (\ref{trrho2}) allows one to rewrite this
inequality as
\begin{equation}
\tr(\bro^{2})-f\geq
\sum_{\mu=1}^{M}\sum_{j=1}^{N_{\mu}}\frac{\tr(\pq_{\mu,j}\bro)^{2}}{a_{\mu}^{2}(b_{\mu}-c_{\mu})}+
\sum_{\mu=1}^{M}\frac{\gamma_{\mu}^{2}(c_{\mu}-f)}{a_{\mu}^{2}(b_{\mu}-c_{\mu})^{2}}
=\sum_{\mu=1}^{M}\sum_{j=1}^{N_{\mu}}\frac{\tr(\pq_{\mu,j}\bro)^{2}}{a_{\mu}^{2}(b_{\mu}-c_{\mu})}
-\frac{1}{d}\,\sum_{\mu=1}^{M}\frac{\gamma_{\mu}}{a_{\mu}(b_{\mu}-c_{\mu})}
\ . \label{fisec}
\end{equation}
Here, we omitted some details based on the use of (\ref{coels}). It
is clear that (\ref{fisec}) is equivalent to the aim (\ref{emt11}).
$\blacksquare$

The inequality (\ref{emt1}) allows us to estimate the index of
coincidence averaged with the weights (\ref{ommu}). In special
cases, it reduces to the results obtained previously, including a
set of MUBs \cite{molm09}, MUMs \cite{rastopn} and $(N,M)$-POVMs
\cite{sent}. In all these cases, the weights (\ref{ommu}) turned out
to be equal. We refrain from presenting the details here.

\end{document}